# Giant Inverse Faraday Effect in Plasmonic Crystal Ring

G. R. Aizin[1†], J. Mikalopas[1], and M. Shur[2*]
[1]*Kingsborough College, The City University of New York, Brooklyn, NY 11235, USA*
[2]*Rensselaer Polytechnic Institute, Troy, NY 12180, USA*

ABSTRACT

Circularly polarized electromagnetic wave impinging on a conducting ring generates a circulating DC plasmonic current resulting in an Inverse Faraday Effect in nanorings. We show that a large ring with periodically modulated width on a nanoscale, smaller or comparable with the plasmonic mean free path, supports plasmon energy bands. When a circularly polarized radiation impinges on such a plasmonic ring, it produces resonant DC plasmonic current on a macro scale resulting in a Giant Inverse Faraday Effect. The metamaterials comprised of the concentric variable width rings ("plasmonic disks") and stacked plasmonic disks ("plasmonic solenoids") amplify the generated constant magnetic field by orders of magnitude.

Ultra-short feature sizes of modern transistors with two-dimensional (2D) electron channel have resulted in ballistic [1,2] or quasi-ballistic [3] operating regimes, where the electron inertia and oscillations of the electron density (i.e., the plasma waves [4-7]) play a dominant role. As consequence, Si CMOS with feature sizes below 30 nm or even larger enabled response [8-11], amplification [12-14] and generation [15-17] at sub-terahertz and terahertz (THz) frequencies. The one bottleneck that emerged is large parasitic effects related to contacts and interconnects that determine the speed of the modern computers [18,19]. This brought attention to plasmonic crystals [20-22] with the unit cell length $l$ being much smaller than the plasmonic mean free path $l_p \sim v_p \tau$, but with the size of the crystal being much larger than $l_p$ and comparable with or even exceeding the wavelength of the THz electromagnetic (EM) radiation. Here $v_p$ is the plasma velocity and $\tau$ is the momentum relaxation time. ($l_p$ is larger than the electron mean free path $l_e = v_F \tau$, where $v_F$ is the electron Fermi velocity.) Plasmonic crystals allow for a decrease in a number of interconnecting circuit elements. [23,24] The plasmonic crystals have been proposed for the generation of sub-THz or THz EM radiation using the structures with the grating gate [25-27] and/or modulated channel width [28,29]. Such plasmonic crystals might have unstable plasmonic modes, both due to the asymmetric reflection from the unit cell boundaries [4,5,25,29] and/or due to the "plasmonic boom" [28] when the electron velocity exceeds the plasma velocity in some fraction of the unit cells [27,28,30].

The ultimate way to eliminate the contacts and interconnects is using ring structures [31] coupled to a circularly polarized EM radiation. In such rings, the plasmons slowly circulate in the direction determined by the helicity of the EM wave generating a circulating DC plasmonic current and constant magnetic field i.e., exhibiting the Inverse Faraday Effect (IFE) [32,33]. Theoretical studies of the IFE in nanorings [31,34,35] revealed the Giant (enhanced by a plasmonic resonances) IFE (GIFE) with the Fano type resonances [34,35]. The estimates in Ref. [35] predicted the magnetic fields of the order of a fraction of Gauss excited by sub-THz or THz beams. However, the area of such plasmonic nanoring is much smaller than a typical waist of a sub-THz or THz beam. Hence, only a small fraction of the THz radiation is converted into the circular current producing the magnetic field. To certain extent, that could be remedied in the structures exciting the "twisted" plasmons in the metamaterials comprising periodic arrays of nanorings or nanoparticles [35]. However, a much stronger effect could be observed in a circular plasmonic crystal, which is a large ring with periodically modulated width on a nanoscale (see Fig. 1a). We demonstrate that such a structure maintains a new type of plasmon bands with spatially varying wavelengths and supports the resonant DC plasmonic current on a macro scale avoiding contacts and interconnects. The metamaterials comprised of the concentric variable width rings (i.e., radially modulated plasmonic structures) significantly amplify the response, since the resonant peaks of the plasmonic ring crystals depend solely on the unit cell structure. Therefore, the plasmonic crystal rings with different radii comprising the plasmonic crystal disk and having the same unit cell have almost the same resonance frequencies (see Fig. 1b). This should lead to a large enhancement in the helicity induced magnetic field.

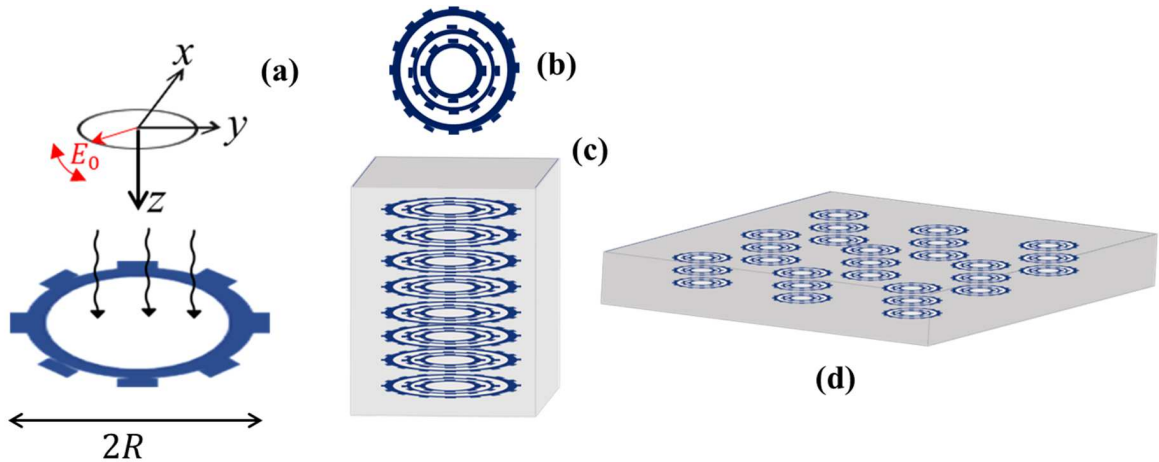

Figure 1. (a) Macroscopic ring with the nanoscale width modulation; (b) Concentric variable width rings ("plasmonic disk"); (c) Stacked periodic array of the concentric variable width rings ("plasmonic solenoid"); (d) Two- dimensional periodic array of the "plasmonic solenoids"

Even larger enhancement can be achieved arranging the concentric ring structures in a stacked periodic array ("plasmonic solenoid") and 2D solenoid array (see Figs. 1c and 1d)

The nonlinearity in the field effect rings enables the detection, mixing, and frequency conversion of the impinging radiation. These plasmonic crystal structures are fabrication compatible with more conventional TeraFETs and TeraFET plasmonic crystals. This extends their applications to the THz spectroscopy [36], frequency-to-digital conversion [37], line-of-sight-detection [38] and other circuits and systems for THz communication and 6G applications (including WIFI 6G).

Below we consider circular plasmonic crystals of variable width (see Fig. 1a) with a unit cell comprising of the two regions of the same length but different width though our analysis applies to any unit cell geometry if its size remains smaller than $l_p$. Our estimates for several semiconductor materials predict the magnetic fields on the order of hundreds of Gausses at reasonable intensities of the impinging helical terahertz radiation.

The linearized hydrodynamic equations for fluctuations of the 2D electron density $\delta n(x,t)$ and hydrodynamic velocity $\delta v(x,t)$ in a ring of radius $R$ with gated electron channel irradiated by a circularly polarized EM wave of frequency $\omega$ and amplitude $E_0$ at normal incidence are

$$\frac{\partial \delta v}{\partial t} = \frac{e}{m^*}\frac{\partial V}{\partial x} - \frac{\delta v}{\tau} \quad , \tag{1}$$

$$\frac{\partial \delta n}{\partial t} + n_0 \frac{\partial \delta v}{\partial x} = 0 \; . \tag{2}$$

Here $x$ is the coordinate along the ring ($0 \leq x \leq 2\pi R$), $n_0$ is the equilibrium 2D electron density, $V(x,t)$ is the electric potential in the 2D channel, and -$e$ and $m$ are the electron charge and effective mass, respectively. Electric potential $V(x,t)$ is the sum of the external potential due to the EM radiation and the potential induced in the channel by the charged electron density fluctuations $e\delta n$. In the gradual channel approximation

$$V(x,t) = -\frac{e\delta n(x,t)}{C} - u_0 e^{-i\kappa x + i\omega t} \tag{3}$$

where $\kappa = \pm \frac{1}{R}$, $u_0 = -\frac{iE_0}{\kappa}$, and signs $\pm$ correspond to two different helicities of the incident EM wave. Here $C = \frac{\varepsilon \varepsilon_0}{d}$ is the gate-to-channel capacitance per unit area, $d$ is the distance between the gate and the channel, and $\varepsilon$ is the dielectric constant of the surrounding medium.

Combining Eq. (3) with Eqs. (1) and (2) one can rewrite the hydrodynamic equations (1) and (2) in the more convenient equivalent form of the telegrapher's equations for the voltage $V$ and the total current $I = -en_0 \delta v W$ in the channel of width $W$ [39]:

$$\frac{\partial V(x,t)}{\partial x} = -\mathcal{L}\frac{\partial I(x,t)}{\partial t} - \mathcal{R}I(x,t) \tag{4}$$

$$\frac{\partial I(x,t)}{\partial x} = -\mathcal{C}\frac{\partial V(x,t)}{\partial t} - i\omega \mathcal{C} u_0 e^{-i\kappa x + i\omega t} \tag{5}$$

Here $\mathcal{C} = CW$, $\mathcal{L} = \frac{m^*}{e^2 n_0 W}$, and $\mathcal{R} = \frac{\mathcal{L}}{\tau}$. Equations (4) and (5) describe the effective plasmonic signal propagating in the transmission line (TL) with distributed capacitance $\mathcal{C}$, inductance $\mathcal{L}$, resistance $\mathcal{R}$, all per unit length, and with a distributed AC voltage source of frequency $\omega$ and amplitude $u_0$ modulated along the TL with the wave number $\kappa$ (see equivalent electric circuit diagram in Fig. 2a.)

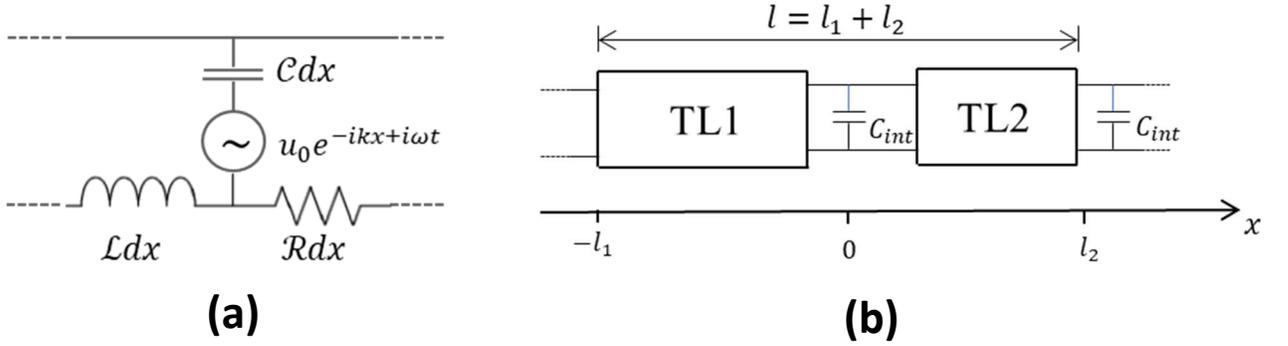

Figure 2. (a) Electric circuit diagram of the equivalent TL representing 2D electron channel in the ring of constant width irradiated by the circularly polarized EM wave at normal incidence; (b) Equivalent electric circuit diagram for the same ring but with periodically changing width, TL1 and TL2 correspond to two segments with different width, $C_{int}$ is the interface capacitance, $l$ is the length of the elementary cell.

The ring with periodically changing width represents a plasmonic crystal with elementary cell of length $l \ll R$ consisting of two segments of widths $W_1$ and $W_2$ and lengths $l_1$ and $l_2$, respectively ($l_1 + l_2 = l$). The difference in widths of the individual segments, $W_1 > W_2$, results in the appearance of the additional displacement currents between the channel and the gate at the segment interfaces
. This effect is accounted for by the interface capacitance $C_{int}$ between the channel and the gate at the segment boundaries: $C_{int} = \alpha C l (W_1 - W_2)$ where constant $\alpha \sim 1$. Figure 2b shows an equivalent electric circuit diagram of the elementary cell in the plasmonic crystal ring. The transmission lines TL1 and TL2 correspond to segments 1 and 2. To analyze the plasmon dynamics in the ring of the variable width, Eqs, (4) and (5) should be solved independently in each segment and supplemented by the boundary conditions at the interfaces between the segments. In the coordinate system shown in Fig. 2b the boundary conditions for the Fourier harmonics of the voltages $V_i(x, \omega)$ and the currents $I_i(x, \omega)$ in each segment ($i = 1, 2$) are

$$V_1(0, \omega) = V_2(0, \omega) \tag{6}$$
$$I_1(0, \omega) = I_2(0, \omega) + \frac{V_1(0,\omega)}{Z_{int}^{(+)}} \tag{7}$$
$$V_2(l_2, \omega) = e^{-ikl} V_1(-l_1, \omega) \tag{8}$$
$$I_2(l_2, \omega) = e^{-ikl} I_1(-l_1, \omega) + \frac{V_2(l_2,\omega)}{Z_{int}^{(-)}} \tag{9}$$

where $\frac{1}{Z_{int}^{(\pm)}} = \pm i\omega C_{int} = \pm i\omega \alpha C l (W_1 - W_2)$, and signs $\pm$ account for the opposite directions of the interface displacement currents at opposite ends of each segment. These boundary conditions represent continuity of the voltage and conservation of the total current at the interfaces between the segments at $x = 0$ and $x = l_2$ in Fig. 2b. In Eqs. (8) and (9), we used the Bloch condition for the currents and voltages in the plasmonic crystal

$$V(x + l, \omega) = e^{-ikl}V(x, \omega) \, , \quad I(x + l, \omega) = e^{-ikl}I(x, \omega). \tag{10}$$

Here $k \in [-\frac{\pi}{l}, \frac{\pi}{l}]$ is the plasmon Bloch wave vector. (In the ring geometry $k = \frac{m}{R}$, $m = \pm 1, \pm 2, \ldots \pm \frac{\pi R}{l}$, $R \gg l$)

The general solution of the telegrapher's equations (4) and (5) for the Fourier harmonics of $V$ and $I$ is

$$\begin{cases} V_i(x, \omega) = A_i e^{-iqx} + B_i e^{iqx} + \frac{q^2 u_0}{\kappa^2 - q^2} e^{-i\kappa x} \\ I_i(x, \omega) = \frac{\omega C W_i}{q}\left(A_i e^{-iqx} - B_i e^{iqx}\right) + \frac{\kappa \omega C u_0 W_i}{\kappa^2 - q^2} e^{-i\kappa x} \end{cases} \quad i = 1,2 \tag{11}$$

where

$$q = -i\sqrt{i\omega \mathcal{C}(\mathcal{R} + i\omega \mathcal{L})} = \frac{1}{v_p}\sqrt{\omega\left(\omega - \frac{i}{\tau}\right)} \tag{12}$$

and $v_p = \sqrt{\frac{e^2 n_0}{m^* \mathcal{C}}}$ is the plasma velocity in the gated 2D electron gas. Constant coefficients $A_i$, $B_i$ are determined by the boundary conditions in Eqs. (6)-(9). In the absence of the external THz radiation ($u_0 = 0$), using Eq. (11) in Eqs. (6)-(9) we obtain the dispersion equation for the plasmon band energy spectrum in the plasmonic crystal formed in the ring with periodically modulated width [40]

$$\cos kl = \cos ql_1 \cos ql_2 - \frac{1}{2}\left(\frac{W_1}{W_2} + \frac{W_2}{W_1} + \frac{\alpha^2 q^2 l^2 (W_1 - W_2)^2}{W_1 W_2}\right) \sin ql_1 \sin ql_2 \tag{13}$$

At $l_1 = l_2 = l/2$ the numerical solution of Eq. (13) with negligible damping ($\tau \to \infty$) yields the plasmonic band spectrum shown in Fig. 3a for $\frac{W_2}{W_1} = 0.4$ and $\alpha = 1$. At small values of $k$, Eq. (13) yields a linear dispersion relation for the plasma waves

$$\omega_{pl,n} = \frac{2\pi v_p}{l} n \pm v_{pl,n}|k| \, , n = 0,1,2, \ldots \tag{14}$$

Here

$$v_{pl,n} = \frac{v_p}{\sqrt{\frac{1}{2} + \frac{1}{4}\left(\frac{W_1}{W_2} + \frac{W_2}{W_1}\right) + \frac{\pi^2 \alpha^2 (W_1 - W_2)^2 n^2}{W_1 W_2}}} \tag{15}$$

is the group velocity of the band plasmon near the center of the Brillouin zone, and $n$ is the band index (see Fig. 3a). The condition $l_p \gg l$ is equivalent to the condition $\omega_{pl,n} \tau \gg 1$ necessary for the resonant plasmonic response.

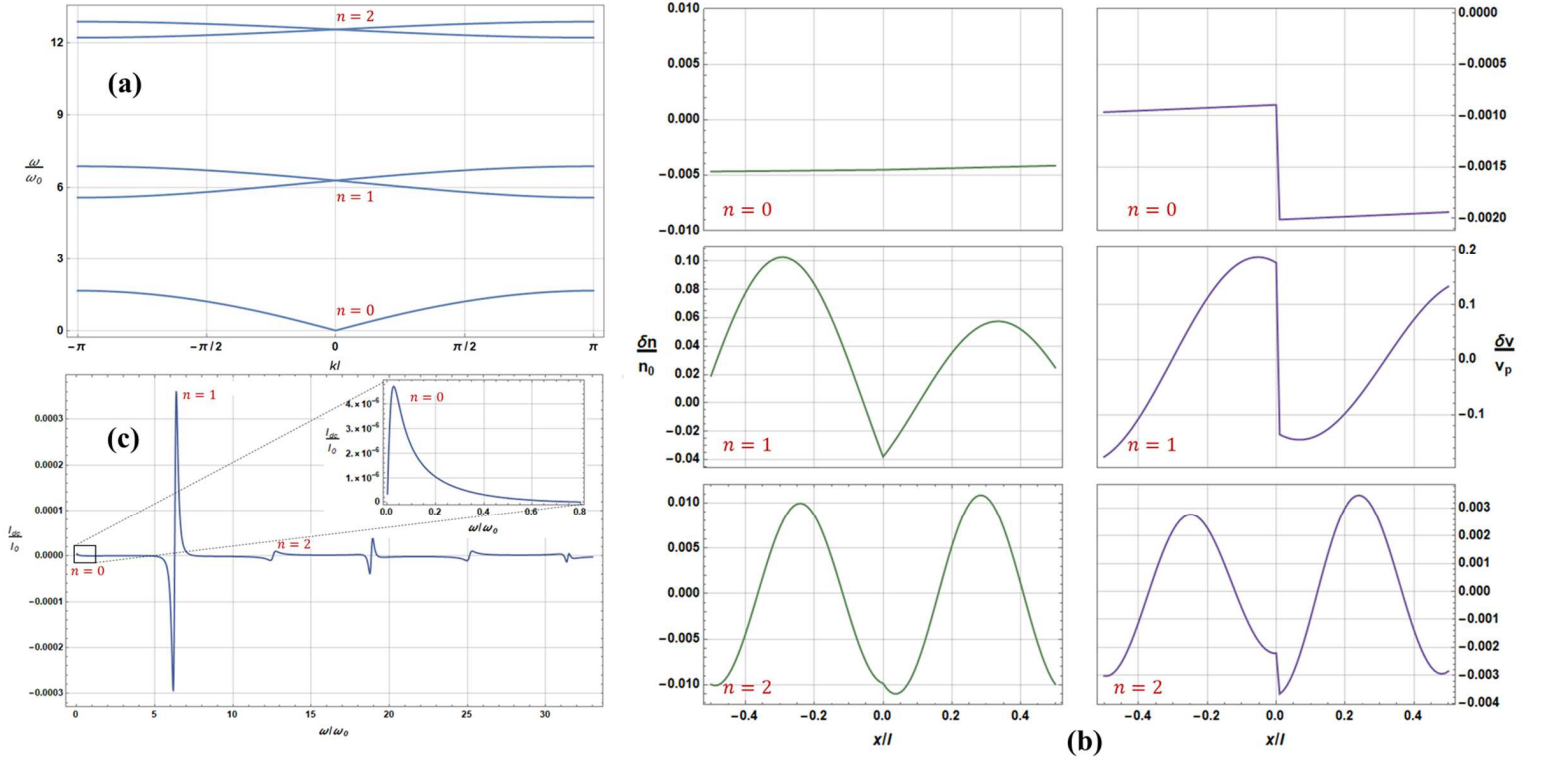

Figure 3. (a) Plasmonic band spectrum in the ring with periodic width modulation, $n$ is the band index; (b) Distribution of the electron density $\delta n$ and the hydrodynamic velocity $\delta v$ in the crystal elementary cell at the frequencies corresponding to the fundamental plasmonic resonance ($n = 0$) and the first two plasmonic band resonances ($n = 1,2$) with positive current peaks; (c) DC current in the circular plasmonic crystal as a function of the frequency of the impinging circularly polarized EM wave. Here $\omega_0 = \frac{v_p}{l}$ and $I_0 = en_0 v_p W_1$.

Impinging circularly polarized THz EM wave of frequency $\omega$ induces traveling electric potential wave in the channel with the same frequency and the wave vector $\kappa = \pm 1/R$ where the direction of propagation is determined by the helicity, see Eq. (3). In the TL formalism, this traveling potential wave is described by the distributed AC voltage source in Eq. (5): $u_0 e^{-i\kappa x + i\omega t}$. This wave resonantly couples to the band plasmons with the Bloch wave vector $\kappa$. To evaluate this effect, we substituted the general solution given by Eq. (11) into Eqs. (6)-(9) with the Bloch wave vector $\kappa$ and solved numerically the resulting system of linear algebraic equations for coefficients $A_i$, $B_i$ in Eq. (11). This solution yielded the spatial distribution of $\delta n(x, \omega)$ and $\delta v(x, \omega)$ in the elementary cell shown in Fig. 3b. It was also used to find the second order DC current resulting from the rectification of the plasma oscillations: $I_{dc} = -e\langle \delta n \delta v \rangle W$ where $\langle ... \rangle$ stands for the averaging over the THz period, see Fig. 3c [40].

Fig. 3c shows the results of the numerical solution for the current $I_{dc}$ as a function of $\omega$ for $\kappa = 1/R$. The dimensionless units used are $\frac{W_2}{W_1} = 0.4$, $l/R = 0.1$, $\tau v_p/l = 3$, $\alpha = 1$, and $|u_0/V_{th}| = 0.01$ where $V_{th} = \frac{en_0}{C}$ is the 2D channel threshold voltage. If the helicity of the THz EM wave is reversed so that $\kappa = -1/R$ the current in Fig. 3c changes sign: $I_{dc}(\kappa = -1/R) = -I_{dc}(\kappa = 1/R)$. As seen in Fig. 3c, all the peaks, with the exception of the fundamental ($n = 0$) peak, have a highly asymmetrical shape (the Fano resonances). The negative and positive peaks correspond to the plasmon spectrum branches degenerate at $k = 0$. For finite $k$, the plasmons forming a standing wave at $k = 0$ split since the finite $k$ breaks the symmetry. The group velocity becomes negative for one of the branches and positive for the other branch. The values of $k$ corresponding to the peaks are equal to $1/R$ and are very close to zero in the scale of Fig. 3a. Therefore, the two adjacent peaks in Fig. 3c are very close. The positive and negative peaks in the induced DC current correspond to the plasmons with the positive and negative group velocity, respectfully (changing signs with the change in the radiation helicity).

Analytical expressions for the current $I_{dc}$ can be derived at the values of frequency $\omega$ close to the plasma resonant frequencies in Eq. (14) in the lowest order on the small parameter $\kappa l \ll 1$. Details of these calculations are described in the Supplemental Material [40]. For the fundamental mode ($n = 0$) we obtain

$$I_{dc} = \frac{en_0}{2v_p}\left(\frac{eE_0}{m^*}\right)^2 \frac{sgn\,\kappa}{(\omega-v_{pl,0}|\kappa|)^2 + \frac{1}{4\tau^2}} \frac{(W_1 W_2)^{3/2}}{(W_1+W_2)^2} \tag{16}$$

For a particular case of $W_1 = W_2$ this expression reduces to that obtained in Ref. [34]. Higher order resonances strongly depend on the parity of the band index $n$:

$$I_{dc} = \pm\frac{en_0}{2v_p}\left(\frac{eE_0}{m^*}\right)^2 \frac{sgn\,\kappa}{\left(\omega-\frac{2\pi v_p n}{l}\mp v_{pl,n}|\kappa|\right)^2 + \frac{1}{4\tau^2}}(W_1+W_2)\begin{cases}\frac{2\pi^2\alpha^2 n^2}{\kappa^2 l^2}\frac{v_{pl,n}^3}{v_p^3}\frac{(W_1-W_2)^2}{W_1 W_2}, & n=1,3,5,\ldots \\ \pi^4\alpha^4 n^4 \frac{v_{pl,n}^5}{v_p^5}\frac{(W_1-W_2)^4}{8W_1^2 W_2^2}, & n=2,4,6,\ldots\end{cases} \tag{17}$$

When $W_1 \to W_2$ the higher order current peaks disappear as expected. It follows from Eqs. (15) and (17) that the height of the resonant peaks decreases when $n$ increases with the height of even peaks being much smaller than that of the odd peaks. These conclusions are confirmed in Fig. 3c.

The most remarkable feature of the rectified DC plasmonic current shown in Fig. 3c is an unusually large strength of the higher order resonant peaks in comparison with the fundamental mode in a similar ring of constant width. The signal enhancement occurs due to the spatial modulation of the electron density in the elementary cells when the plasmonic band modes are excited. The modulation induces electric field of the order $u_0/l$. This induced electric field is much stronger than the external electric field of the order $u_0/R$, which determines the strength of the fundamental peak. Fig. 3b shows the plots of the spatial distribution of the electron density $\delta n(x)$ and the hydrodynamic velocity $\delta v(x)$ in the crystal elementary cell at resonant frequencies corresponding to the positive DC current peaks with $n = 0,1,2$ confirming our qualitative conclusions.

The DC current in the rings irradiated by the circularly polarized EM wave produces constant magnetic field. This magnetic field recently estimated for the nanorings of the constant width [35] be on the order of 0.1 Gs. Using the rings of variable width and plasmonic crystals significantly expands the range of the achieved magnetic fields. Table I lists the estimates of the GIFE-induced magnetic fields in the first plasmonic band resonance ($n = 1$) for the two TeraFET ring systems implemented using different materials: the concentric rings forming a plasmonic disk and the plasmonic solenoid formed by the stacked plasmonic disks. Material parameters are taken from Ref. [35]. The notation used in Table I is as follows: $f_1$ is the resonant plasma frequency of the 1$^{st}$ band mode in the plasmonic ring crystal, $I_1$ is the DC current in the outermost ring of the plasmonic disk, $B$ is the magnetic field at the center of the plasmonic disk, $B_{sol}$ is the magnetic field in the plasmonic solenoid formed by the stacked separated plasmonic disks. All estimates are done for $W_1/W_2$ =15nm/10nm and $l$ =100nm. The values of $B$ are found for the plasmonic disk with 361 concentric rings with the inner radius of 1µm, outer radius of 10µm and the ring separation of 0.025µm. $B_{sol}$ is estimated for the plasmonic solenoid assuming 1µm separation between the plasmonic disks. The values of $B$ and $B_{sol}$ are estimated using the model developed above. This model is valid for relatively small intensities of the THz radiation when the DC current in the rings is proportional to the intensity and the THz radiation induced variation of the electron density $\delta n$ is much smaller than the equilibrium electron density $n_0$ in the ring. The values of the electric field amplitude $E_0$ of the THz EM radiation and the EM power through the disk $P$ used in this calculation and satisfying the above condition are also listed in the Table I.

At larger THz intensities, the electron current in the rings saturates and $\delta n$ becomes comparable to $n_0$. Hence, the electron saturation current determines the GIFE dynamic range and the "ultimate" magnetic field. The values of $B^{ult}$ are estimated for the plasmonic disk assuming that the THz intensity is high, and the current is saturated in all rings of the plasmonic disk with saturation velocity $\sim 10^5$m/s. Finally, the "ultimate" plasmonic solenoid magnetic field $B_{sol}^{ult}$ is estimated for a plasmonic disk with outer radius of 100µm, the separation between the disks 50nm and assuming the current saturation. Such a solenoid could be fabricated using the technology similar to the All Around Gate (AAG) technology that uses even smaller dimensions [19]. We should also notice that, due to the overshoot effects, the peak electron velocity could greatly exceed the saturation velocity in the pulsed operation [41,42] and, therefore, even larger magnetic field could be achieved.

Table I Estimates of the GIFE-induced magnetic field for the ring-shaped TeraFET systems implemented using different materials systems.

| | $f_1$(THz) | $I_1$ (µA) | $B$ (µT) | $B_{sol}$ (µT) | $E_o$ (V/m) | $P$ (µW) | $B^{ult}$(mT) | $B_{sol}^{ult}$ (mT) |
|---|---|---|---|---|---|---|---|---|
| InGaAs 77K | 15.4 | 0.58 | 11.4 | 136 | 7,200 | 43.6 | 0.279 | 87.2 |
| InGaAs 300K | 15.4 | 0.23 | 4.89 | 55 | 21,600 | 392 | 0.279 | 87.2 |
| GaN 77K | 17.6 | 17.8 | 180 | 2725 | 7,200 | 43.6 | 1.40 | 436 |
| GaN 300K | 17.6 | 0.7 | 15 | 169 | 72,000 | 4,360 | 1.40 | 436 |
| Si 77K | 4.1 | 0.23 | 4.84 | 56 | 2,900 | 7.00 | 0.285 | 88.8 |
| Si 300K | 4.1 | 0.056 | 0.722 | 9.58 | 29,000 | 700 | 0.285 | 88.8 |
| p-diamond 77K | 2.6 | 0.43 | 6.22 | 84.6 | 723 | 0.44 | 0.279 | 87.2 |
| p-diamond 300K | 2.6 | 0.375 | 1.42 | 16 | 4,200 | 15.3 | 0.279 | 87.2 |

Converting electromagnetic field into a constant magnetic field is a classical solid state physics problem with the IFE being one of the most popular solutions. As shown in Ref. [35], plasmonic resonances in TeraFET nanorings could substantially enhance this IFE magnetic field converting this effect to GIFE. We now show a further dramatic increase could be achieved using plasmonic disks and solenoids comprised of the modulated width plasmonic rings. The response can be increased even more by breaking the symmetry of the unit cell. Also, under the short pulse excitations, overshoot, and ballistic response on a femtoscale allows obtaining much larger transient currents and, therefore, much larger magnetic fields. The ungated rings should exhibit similar effects and may be preferable because the ungated plasmonic disks and especially ungated plasmonic solenoids are much easier to fabricate. The assemblies of the plasmonic disks and plasmonic solenoids are the new type of a helical metamaterial suitable not only for generating large magnetic field but also for detection, mixing, and frequency conversion of THz radiation. Future studies should include the investigations of these applications.

The work at RPI and CUNY was supported by the US Army Research Office (Project Manager Dr. Joe Qiu.)

† gaizin@kbcc.cuny.edu
* shurm@rpi.edu

## Supplemental Material

Here, we describe the main steps used to derive the results presented in the main text.

*Plasma spectrum in the plasmonic crystal ring*
Substituting general solution for $V_i(x,\omega)$ and $I_i(x,\omega)$ from Eq. (11) into the boundary conditions given by Eqs. (6)-(9) we obtain the following system of linear inhomogeneous equations for unknown coefficients $A_i, B_i, i = 1,2$:

$$A_1 + B_1 - A_2 - B_2 = 0$$

$$\left(\frac{\omega C W_1}{q} - \frac{1}{Z_{int}^{(+)}}\right)A_1 - \left(\frac{\omega C W_1}{q} + \frac{1}{Z_{int}^{(+)}}\right)B_1 - \frac{\omega C W_2}{q}A_2 + \frac{\omega C W_2}{q}B_2 = \frac{q^2 u_0}{(\kappa^2 - q^2)Z_{int}^{(+)}} + \frac{\kappa \omega C(W_2 - W_1)u_0}{\kappa^2 - q^2}$$

$$e^{-i\kappa l}e^{iql_1}A_1 + e^{-i\kappa l}e^{-iql_1}B_1 - e^{-iql_2}A_2 - e^{iql_2}B_2 = 0 \qquad (S1)$$

$$\frac{\omega C W_1}{q}e^{-i\kappa l}e^{iql_1}A_1 - \frac{\omega C W_1}{q}e^{-i\kappa l}e^{-iql_1}B_1 - \left(\frac{\omega C W_2}{q} - \frac{1}{Z_{int}^{(-)}}\right)e^{-iql_2}A_2 + \left(\frac{\omega C W_2}{q} + \frac{1}{Z_{int}^{(-)}}\right)e^{iql_2}B_2 =$$

$$-\frac{q^2 u_0 e^{-i\kappa l_2}}{(\kappa^2 - q^2)Z_{int}^{(-)}} + \frac{\kappa \omega C(W_2 - W_1)u_0 e^{-i\kappa l_2}}{\kappa^2 - q^2}$$

In the absence of the external EM wave ($u_0 = 0$) the right-hand side of the system of equations (S1) is equal to zero. The requirement of the non-zero solution of the remaining linear homogeneous system of equations yields the plasmon dispersion equation (13) in the main text.

*Forced plasma oscillations in the plasmonic crystal ring*
To describe forced plasma oscillations caused by an external circularly polarized EM wave of frequency $\omega$ the system of equations (S1) was numerically solved for unknown coefficients $A_i, B_i$ as a function of $\omega$. These solutions were used in Eq. (11) of the main text to find spatial distributions of the voltage $V_i(x,\omega)$ and the current $I_i(x,\omega)$ in each segment ($i = 1,2$) of the crystal elementary cell and then to find spatial distributions of the electron density $\delta n(x,\omega)$ and hydrodynamic velocity $\delta v(x,\omega)$ defined as

$$\delta n_i(x,\omega) = -\frac{C}{e}\left(V_i(x,\omega) + u_0 e^{-i\kappa x}\right),$$

$$\delta v_i(x,\omega) = -\frac{I_i(x,\omega)}{e n_0 W_i}. \qquad i = 1,2 \qquad (S2)$$

Examples of these distributions for $Re\delta n(x,\omega)$ and $Re\delta v(x,\omega)$ at several resonant frequencies are shown in Fig. 3b.
The DC plasmonic current $I_{dc}(\omega)$ does not depend on the coordinate $x$ along the ring and can be found by performing the time averaging over the THz period at any point $x$ as

$$I_{dc}(\omega) = -\frac{e}{2}Re\big(\delta n(x,\omega)\delta v^*(x,\omega)\big)W(x) \qquad (S3)$$

where $W(x)$ is the width of the ring at point $x$. In the next section, this is confirmed by the direct analytical calculation of the current at the frequencies close to the resonant plasma frequencies in Eq. (14) of the main text. However, for computational convenience we used $x$-averaging to find $I_{dc}(\omega)$ as

$$I_{dc}(\omega) = -\frac{eW_1}{2l}\int_{-l_1}^{0} Re\big(\delta n(x,\omega)\delta v^*(x,\omega)\big)dx - \frac{eW_2}{2l}\int_{0}^{l_2} Re\big(\delta n(x,\omega)\delta v^*(x,\omega)\big)dx \qquad (S4)$$

The result is shown in Fig. 3c at $l_1 = l_2 = l/2$.

*DC plasmonic current near the plasma resonances in the plasmonic crystal ring*

Analytical expressions for the DC plasmonic current can be derived for the frequencies of the external EM radiation close to the resonant plasma frequencies in Eq. (14) of the main text as described below.

Coefficients in the system (S1) were expanded into a series using the small parameters $\kappa l \ll 1$ and $\delta\omega l/v_p \ll 1$ where $\delta\omega = \omega - \frac{2\pi n v_p}{l}, n = 0,1,2 \ldots$ is the deviation of the external frequency from the resonant plasma frequency at $k = 0$. We retained the lowest order non-vanishing terms in this expansion and solved the resulting simplified system of algebraic equations for unknown coefficients $A_i$ and $B_i$. These solutions were used in Eq. (11) of the main text to find distribution of the voltages $V_i(x,\omega)$ and the currents $I_i(x,\omega)$ in the segments of the elementary cell of the plasmonic crystal. Substituting these distributions into Eqs. (S2) we obtained expressions for the fluctuations of the electron density $\delta n_i(x,\omega)$ and the hydrodynamic velocity $\delta v_i(x,\omega)$ in each segment near plasmonic resonances $\omega_{pl,n}$ in the following general form

$$\delta n_i(x,\omega) = -\frac{C}{e}\left[\alpha_i(\omega)\cos\frac{2\pi n x}{l} + i\beta_i(\omega)\sin\frac{2\pi n x}{l}\right]$$
$$\delta v_i(x,\omega) = \frac{Cv_p}{en_0}\left[\beta_i(\omega)\cos\frac{2\pi n x}{l} + i\alpha_i(\omega)\sin\frac{2\pi n x}{l}\right] \qquad (S5)$$

Complex coefficients $\alpha_i$ and $\beta_i$ depend on the parity of the band index *n* and have resonant behavior near the plasma frequencies given by Eq. (14). We do not write out explicit expressions for these coefficients because of the very cumbersome form of the derived formulas

The DC plasmonic current was found in each segment using Eq. (S5) as

$$I_{dc}^{(1,2)}(\omega) = -eW_{1,2}\frac{1}{2}Re\left[\delta n_{1,2}\delta v_{1,2}^*\right] = \frac{C^2 W_{1,2} v_p}{2en_0}\left[\alpha_{1,2}'(\omega)\beta_{1,2}'(\omega) + \alpha_{1,2}''(\omega)\beta_{1,2}''(\omega)\right] \qquad (S6)$$

This current does not depend on the coordinate *x* as expected. Calculations also show that $I_{dc}^{(1)} = I_{dc}^{(2)}$. The results of our analytical calculations of the DC plasmonic current are presented in Eqs. (16) and (17) of the main text.